\documentclass[12pt,preprint]{aastex}
\usepackage{psfig,natbib}

\def\simlt{\mathrel{\hbox{\rlap{\hbox{\lower4pt\hbox{$\sim$}}}\hbox{$<$}}}}
\def\simgt{\mathrel{\hbox{\rlap{\hbox{\lower4pt\hbox{$\sim$}}}\hbox{$>$}}}}

\def\ale{\mathrel{\hbox{\rlap{\hbox{\lower4pt\hbox{$\sim$}}}\hbox{$<$}}}}
\def\age{\mathrel{\hbox{\rlap{\hbox{\lower4pt\hbox{$\sim$}}}\hbox{$>$}}}}

\def\spose#1{\hbox to 0pt{#1\hss}}
\newcommand\lsim{\mathrel{\spose{\lower 3pt\hbox{$\mathchar"218$}}
     \raise 2.0pt\hbox{$\mathchar"13C$}}}
\newcommand\gsim{\mathrel{\spose{\lower 3pt\hbox{$\mathchar"218$}}
     \raise 2.0pt\hbox{$\mathchar"13E$}}}

\begin{document}


\title{Direct Detection of Dark Matter with Space-based Laser Interferometers}

\author{A. W. Adams\altaffilmark{1}, J. S. Bloom\altaffilmark{1,2}}

\affil{$^1$ Harvard Society of Fellows, 78 Mount Auburn Street, Cambridge, 
MA 02138 USA}

\affil{$^2$ Harvard-Smithsonian Center for Astrophysics, MC 20,
60 Garden Street, Cambridge, MA 02138, USA}


\begin{abstract}

Dark matter pervades the Solar System, free-streaming at the local
Galactic orbital velocity of the halo with a space density of $\sim 9
\times 10^{-25}$\,gm cm$^{-3}$. As these objects pass through the
Solar system, they perturb gravitationally, and thus very weakly, all
nearby inertial masses.  Making use of this, we propose an approach to
the direct detection of dark matter at previously inaccessible
intermediate masses ($10^{14} - 10^{20}$\,gm). Such mass scales are
relevant, for example, for dark matter made of primordial black holes
or clumped matter in a sequestered sector. If such dark matter exists,
it will be unambiguously detectable through its inelastic
gravitational interaction with the proposed Laser Interferometer Space
Antenna (LISA) experiment.  We demonstrate the efficacy of this
approach by studying the dark matter signal in numerical simulations
of the LISA data stream.  A more conservative approach --- to detect
dark matter in the differential acceleration power spectrum ---
significantly underestimates the expected rates for
LISA. Interestingly, while the space-density of 10$^{15}$\,gm DM
objects would be comparable to the space-density of asteroids of
similar masses, such ``light matter'' contaminants are readily
detectable in reflected Solar light, allowing for the elimination of
the major background contaminant.
\end{abstract}



\section{Introduction}

Only a small fraction of the mass in the universe has been directly
detected through its electromagnetic signature; the existence of the
remaining dark matter \citep{zwi33} is inferred from its gravitational
fingerprint.  This dark matter is believed to comprise a
free-streaming gas of massive objects whose detailed nature is largely
unconstrained; viable models posit masses anywhere between a few
electronvolts and millions of solar masses.  Searches on both the high
\citep{aaa+01} and low \citep{hks+98} ends have so far led to no
convincing account for the missing mass in the universe. Dark matter is
the dominant component of the gravitational potential of galaxy
clusters, galaxies, and perhaps galactic disks
\citep{tre92,carr94}. Searches for Galactic halo gravitational
microlensing \citep{pac86} of planet-mass (and larger) dark matter
suggest that no more than $\sim$40\% of the baryonic component of dark
mass in the Galaxy can be contained in such compact objects
(``MACHOs'') with masses from $10^{26}$ -- $10^{34}$ gm
\citep{aaa+01}.  On particle physics scales, searches for weakly
interacting massive particles (WIMPs) have thus far produced only
negative results --- however, these experiments assume weak
non-gravitational couplings to ordinary matter, and so offer no bounds
on purely gravitationally-coupled dark matter. Dark matter on mass
scales less than $10^{25}$\,gm have never been searched for through
gravitational interactions.

From matching observations to theory \citep{bdpr84,defm85,deab+00}, it
is believed that dark matter comprises a free-streaming
(non-interacting) fluid of cold (non-relativistic) compact objects.
The nature of the constituents of this cold free-streaming fluid is
almost completely model dependent. The possible origins of DM at the
high and low mass range have been reviewed elsewhere
\citep{Griest:1995gs,Afshordi:2003zb,Bertone:2004pz}.  At intermediate
masses, there are two broad classes of popular DM candidates. First,
it is possible that a large fraction of this fluid is composed of
microscopic primordial black holes (PBHs) \citep{haw71,ch74,cgl03}; the
mass scale for these is set by the dual requirement that (a) they
comprise a significant fraction of the inferred dark matter energy
density, and (b) were created (at some point well before Big Bang
Nucleosynthesis) at sufficiently high mass that the majority have not
yet Hawking-evaporated \citep{haw75}; this suggests that surviving PBH
have masses, very roughly, somewhere over $10^{15}$\,gm.  Secondly,
the dark matter could be formed of non-baryonic matter typical in
models with extra dimensions, for example fields on a distant brane in
string theory \citep{add}; since the physics of such purely
gravitationally-coupled systems is, by definition, beyond our ability to
constrain from non-gravitational observations, in this paper we take
the maximally agnostic stance that such systems might produce compact
objects on any mass scale, relying on observations to rule them out.

Ontology aside, Galactic halo constituents of any mass are thought to
be isotropically distributed and free-streaming through the Solar
System with Galactic-orbital velocity ($v_{\rm circ} \approx 230$ km
s$^{-1}$), with the dark mass density in the neighborhood of the Solar
System equal to the local Galactic halo density, about $\rho_{{\rm
DM}} = 9 \times 10^{-25}$ gm cm$^{-3} \equiv \rho_{\rm {DM, halo}}$
\citep{ggt95}.  This value of $\rho_{{\rm DM}}$ is consistent with 
upper-limits on increased perihelion precession of Solar System
bodies \citep{gs96} ($\ale 10^{27}$\,gm dark matter mass contained
within the orbit of Uranus).  Note that with this DM density and
characteristic velocity, the Sun would accrete only $\approx 10^{-9}$
$M_{\odot}$ dark matter over its lifetime.

It is important to emphasize the following: if DM objects couple only
gravitationally to ordinary matter, they may well pervade the Solar
System without ever having been noticed.  For example, asteroids of
masses similar to those considered herein are easily detected when
colliding with a planet not because of gravitational interactions, but
due to electromagnetic interactions, which result in the vaporization
of the asteroid and the accompanying catastrophe for the planet. A
truly {\it dark} asteroid, however, would sail through the planet
almost as quietly as a neutrino.

\section{Detecting dark matter with LISA}

A dark matter object streaming through the Solar System will
gravitationally scatter a test mass; the deflection of the orbit of
this test mass is thus a direct measurement of the passage of the dark
object.  Since the profound weakness of gravitational interactions
makes the magnitude of these deflections extremely small, our ability
to detect dark objects via gravitational scattering is limited by our
ability to accurately map the motion of inertial test masses.

The 5-year LISA experiment \citep{ben+98}, due to launch in 2011, will
feature three spacecraft, forming an equilateral triangle of arm
length $b = 5 \times$ 10$^{11}$\,cm, in an Earth-trailing orbit. The
interferometric design calls for an rms positional accuracy of $\sigma
= 10^{-9}$\,cm every sample at a rate of $S = 1$\,Hz. With such
precision, the passage of a sufficiently massive body should be, in
principle, easily seen: taking $M_{\rm DM} = 10^{15}$\,gm $\equiv
M_{15}$, the typical distance to the nearest object is $l_{\rm typ}
\approx 10^{13}$\,cm $M_{15}^{1/3}$ $\rho_{\rm {DM, halo}}^{-1/3}$,
less than the Earth-Sun distance. If $\vec l$ is chosen as the
distance of closest approach (``impact parameter''), then over a
timescale of $|\vec l|/v \approx 5$\,days ($l/l_{\rm typ}$), the dark
object will effectively displace the test mass impulsively in a
direction parallel to $\vec l$ by $(l/v)^2 G M_{15}/2 l^2 \approx 6.3
\times 10^{-8}$\,cm, where $G$ is Newton's constant. This impulsive
displacement over 5 days is $\approx$60 times the rms noise on a
single 1 sec sample from LISA.

In the absence of other forces, and with long-lived coherence ($>$
week) in absolute positioning, this signal would be rather
straightforward to detect.  However, two effects complicate the
detection considerably. First, only the change in the separation
between pairs of LISA stations is measured, so the effect of a passing
DM object will only be seen if the DM object imparts a substantial
{\it differential} (i.e., tidal) force between two stations, that is
when the impact parameter to one station is on the order of the LISA
arm length or smaller.  Secondly, the Keplerian orbits of the
spacecraft and the presence of major Solar System bodies perturb the
inter-station distances by factors of $\sim$10$^{17}$ and
$\sim$10$^{13}$, respectively, over the expected DM signal (see Figure
2).

\subsection{Conservative Analytic Estimate of Rates: Detecting DM in the acceleration power-spectrum}  

A very conservative estimation of the S/N ratio of such measurements
may be obtained by exploiting the well-developed techniques in the
pre-phase A study for LISA \citep{ben+98}, which calculate S/N ratios
in the acceleration power-spectrum of a passing asteroid.  While this
is the correct technique for estimating the background due to
asteroids and comets obscuring the oscillatory signal from
gravitational waves, this is very far from the optimal technique for
detecting impulsive near-field events, as discussed above;
nonetheless, it gives a pleasingly reasonable, if unduly conservative,
measure of the expected number of events.  Using these techniques, we
estimate that a (S/N) $> 3$ would be obtained for an impact parameter
to one spacecraft of $l_{\rm max} < 4.2 \times 10^{10}$\,cm
$M_{15}^{2/3} v_{230}^{-1/3}$, for impacts between $6.1
\times 10^{8} v/v_{230}$\,cm $< l \ale$ $b/3 = 1.7 \times
10^{11}$\,cm. Here the lower limit bounds the analytic approximation
to the S/N integration (at smaller $l$, the value of $l_{\rm max}
\propto M$). The upper limit is imposed since at larger impact, tidal
forces on LISA become important and only the {\it differential}
acceleration can be measured. In this case, for $l_{\rm max} > b/3$ we
estimate the maximum impact parameter to be:
\begin{equation}
l_{\rm max} = \left[\frac{13 \pi \left(G M b\right)^2}{64 (S/N)^2 v
a_c^2}\right]^{1/5},
\end{equation}
where $a_c = 6 \times 10^{-13}$ cm s$^{-2}$/$\sqrt{\rm Hz}$ (see
\citealt{ben+98}). The value of $l_{\rm max} = b/3$ when $M \approx
3.5 M_{15}$. Thus, for passages larger than $b/3$, only masses greater
than $\approx 3.5 M_{15}$ will produce a measurable effect in
acceleration power-spectrum.

To arrive at an approximate upper bound on the expected event rate
$R(M)$ from this analysis, we assume that all DM have the same mass
and that the flux of DM objects is $f(M) = \rho_{\rm {DM, halo}}\,
v_{230}/M = 2.1 \times 10^{-32} M_{15}/M$ cm$^{-2}$ s$^{-1}$. The
value of $R(M) = \pi l_{\rm max}^2 f(M)$ is $6.4 \times 10^{-10}
(M/M_{15})^{1/3}$ s$^{-1}$ for $M \ale 5 M_{15}$ and $4.0\times
10^{-10} (M/10 M_{15})^{-1/5}$ s$^{-1}$ for $M
\age 5 M_{15}$ (that is, the peak sensitivity of LISA using this method is around $M_{\rm DM} = 5 \times 10^{15}$\,gm)
Multiplying this by 3 (for each station) and by 5 years, the expected
number of DM events detected in acceleration-frequency space by LISA
is $\sim$0.04 (for $M_{\rm DM} = M_{15}$). We note that with only a
few orders of magnitude decrease in the system acceleration noise, the
next generation gravitational-wave interferometer should easily detect
several events, if indeed dark matter on these mass scales exists.

\subsection{Detecting the Impulsive Signals from DM: A Simulation}

It is very encouraging that the expected number of events approaches
unity with the conservative measurement of S/N using acceleration
power spectra. However, we emphasize that unlike LISA signals from
weak gravitational waves, the DM passages by LISA are strongly
inelastic, changing the optimal detection scheme in several important
ways.  Most importantly, the fact that the interaction is impulsive
rather than oscillatory means that even when peak acceleration at
nearest approach is relatively small, the net displacement may be
quite large.  Furthermore, since each such inelastic collision changes
the circumference, such perturbations can be unambiguously attributed
to near-field objects rather than gravitational waves (this center
sagnac channel is actually a rejection trigger for LISA's
gravitational-wave mission).  Additionally, since the major-body
near-field interaction timescales are significantly longer than the
timescale for lower-mass DM scattering events, the DM signal may be
explicitly decoupled from major-body perturbations (see Figure 3).
Finally, since the nearest-approaches to the various LISA stations are
separated by $b/v_{\rm DM} \approx 6$ hr time lags, detecting (or
predicting) such duplications in the impulsive signal provides a
powerful check.

Unfortunately, the role of instrumental noise and decorrelation in 
determining a precise S/N measure for this impulsive-search strategy 
will require a new set of techniques (we leave this analysis to future 
work).  As a first test the feasibility of this search strategy, we have run
extensive simulations of the LISA experiment to determine whether 
such DM scattering events are in fact detectable in the LISA datastream.
As will be clear from the figures, these inelastic collisions with the LISA 
constellation are readily seen in the impulsive signal on the inter-station
timeseries on scales significantly larger than LISA's 1 sec rms accuracy.

The simulations essentially keep track of the perturbation from their
Keplerian orbits of the three LISA stations due to all relevant
massive objects in the solar system.  The basic LISA orbits were taken
from the equations presented in \citet{cr03}, and are recorded to
second order in the eccentricity of the LISA constellation. The
"major-body" perturbations are those due to all inner Solar system
planets and the Moon whose positions are determined from their known
orbits.

Smaller perturbations are due to the simulated dark matter with fixed
mass density $\rho_{\rm DM} = 9 \times 10^{-25}$\,gm cm$^{-3}$. We
generate an isotropic distribution of DM masses, choosing random
starting positions of $N$ DM objects in a volume $V$ (set so that $N =
\rho_{\rm DM} \times V / M$). For each DM object, a randomly oriented
velocity is chosen with magnitude drawn from a Maxwellian distribution
(with $v_0= v_{\rm circ} = 230$ km s$^{-1}$) truncated at the escape
velocity of the Galaxy ($v_{\rm esc} = 600$ km s$^{-1}$) at the Solar
distance from the Galactic center \citep{ls96}.  The motion of
individual DM objects is assumed to be free streaming (ie.,
acceleration is zero) through the simulation volume, and the object is
reflected through the origin once reaching the volume edge. The
typical streaming time through the volume is $\sim V^{1/3}/v \approx
30$\, days. The free streaming assumption is a reasonable
approximation since the escape velocity from the Solar potential at 1
AU is 42 km s$^{-1}$ so DM trajectories on length scales $< 1$ AU are
not significantly altered by the stellar nor planetary potentials. The
gravitational coupling between DM objects has a negligible affect on
the trajectories.

\begin{figure*}[tp]
\centerline{\psfig{file=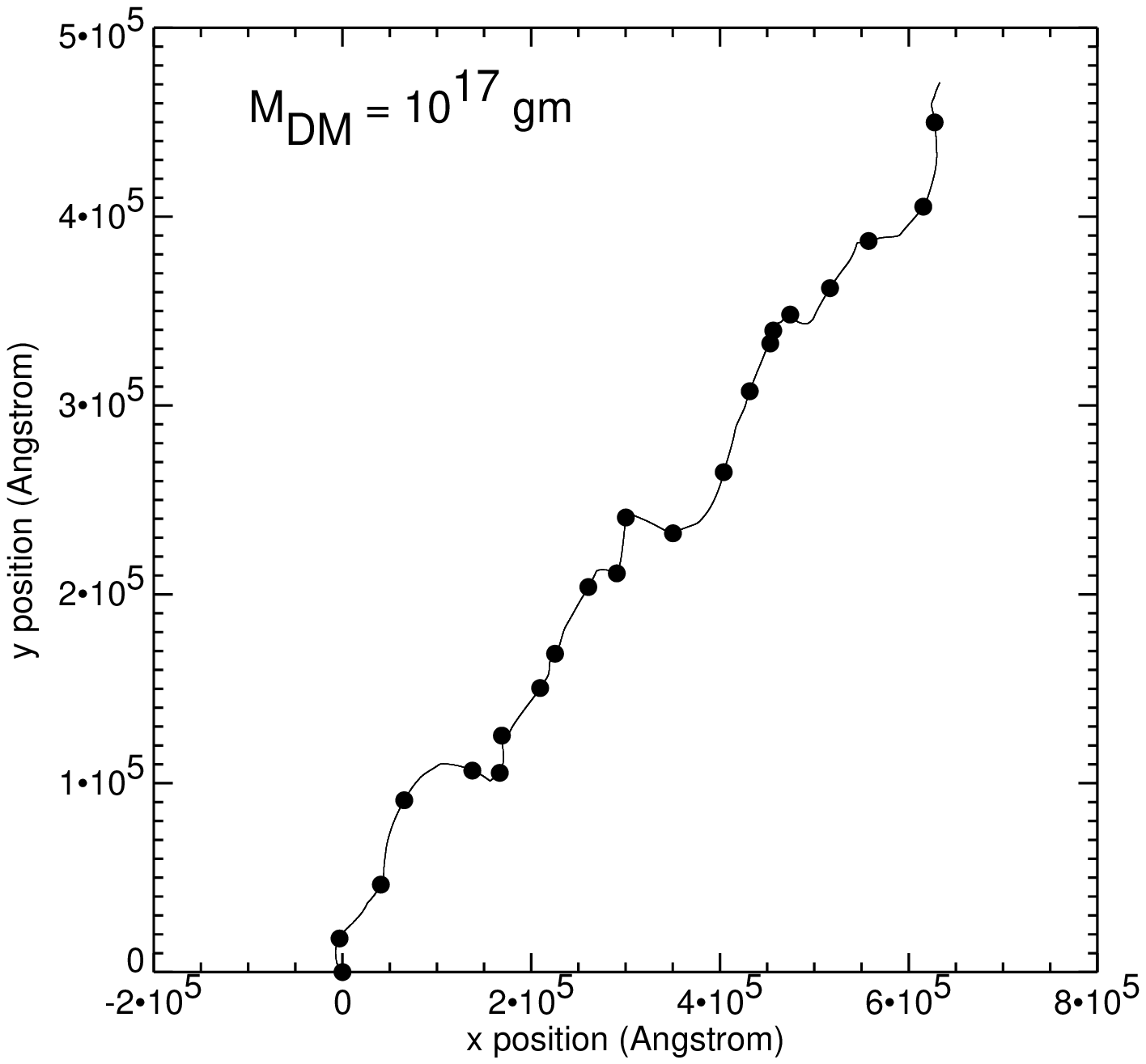,width=3.2in,angle=0}
\psfig{file=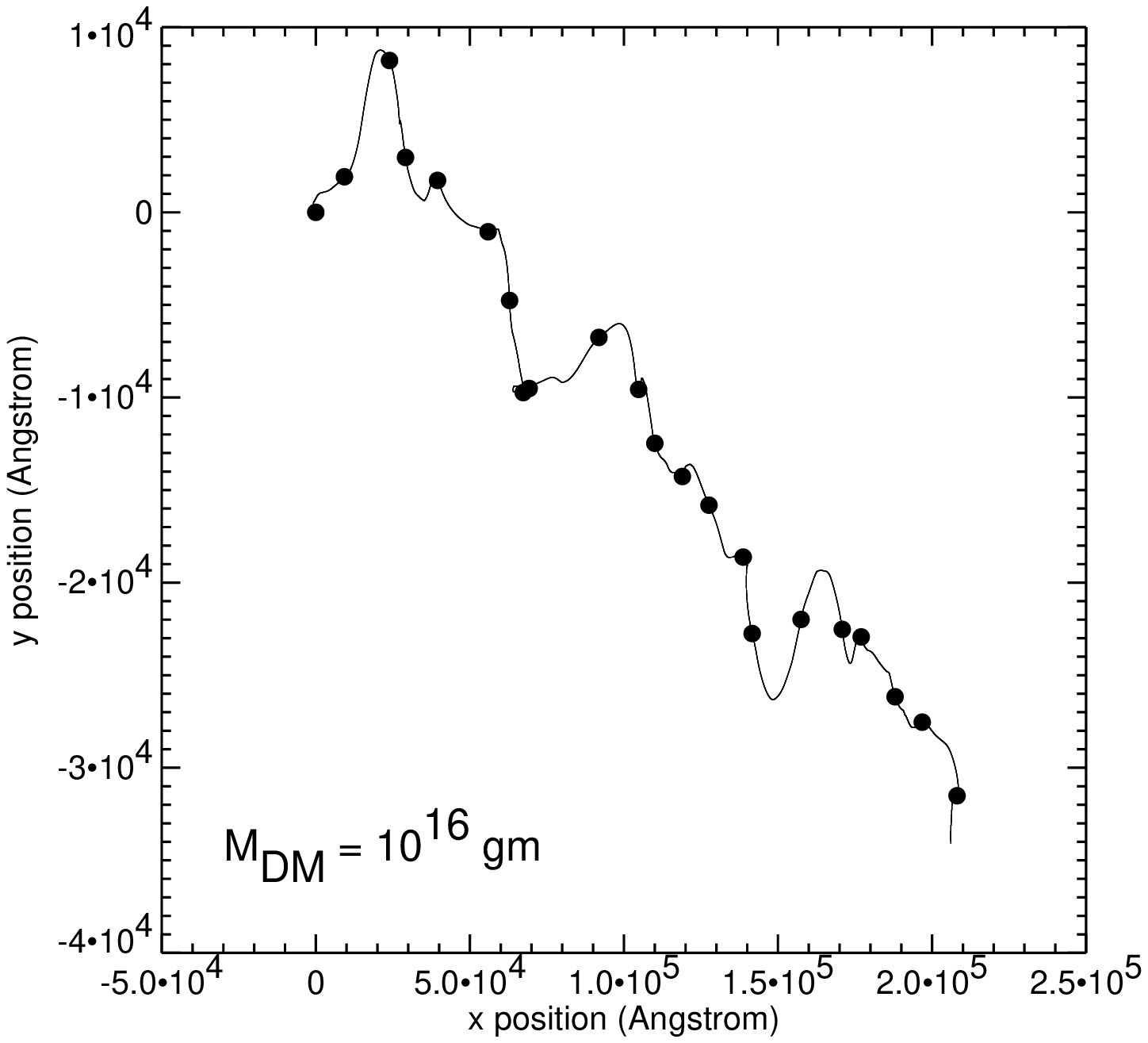,width=3.2in,angle=0}
}
\centerline{\psfig{file=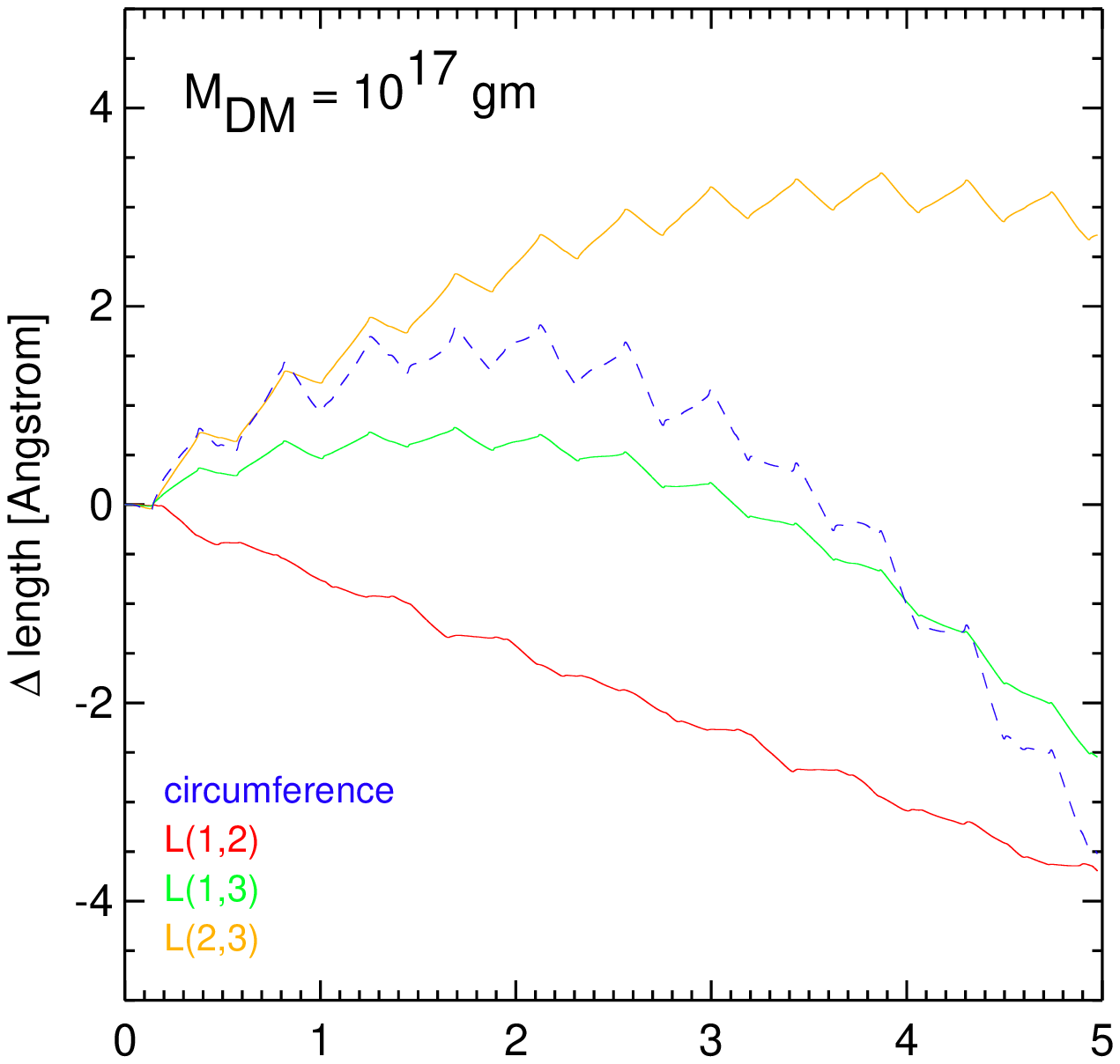,width=3.2in,angle=0}
\psfig{file=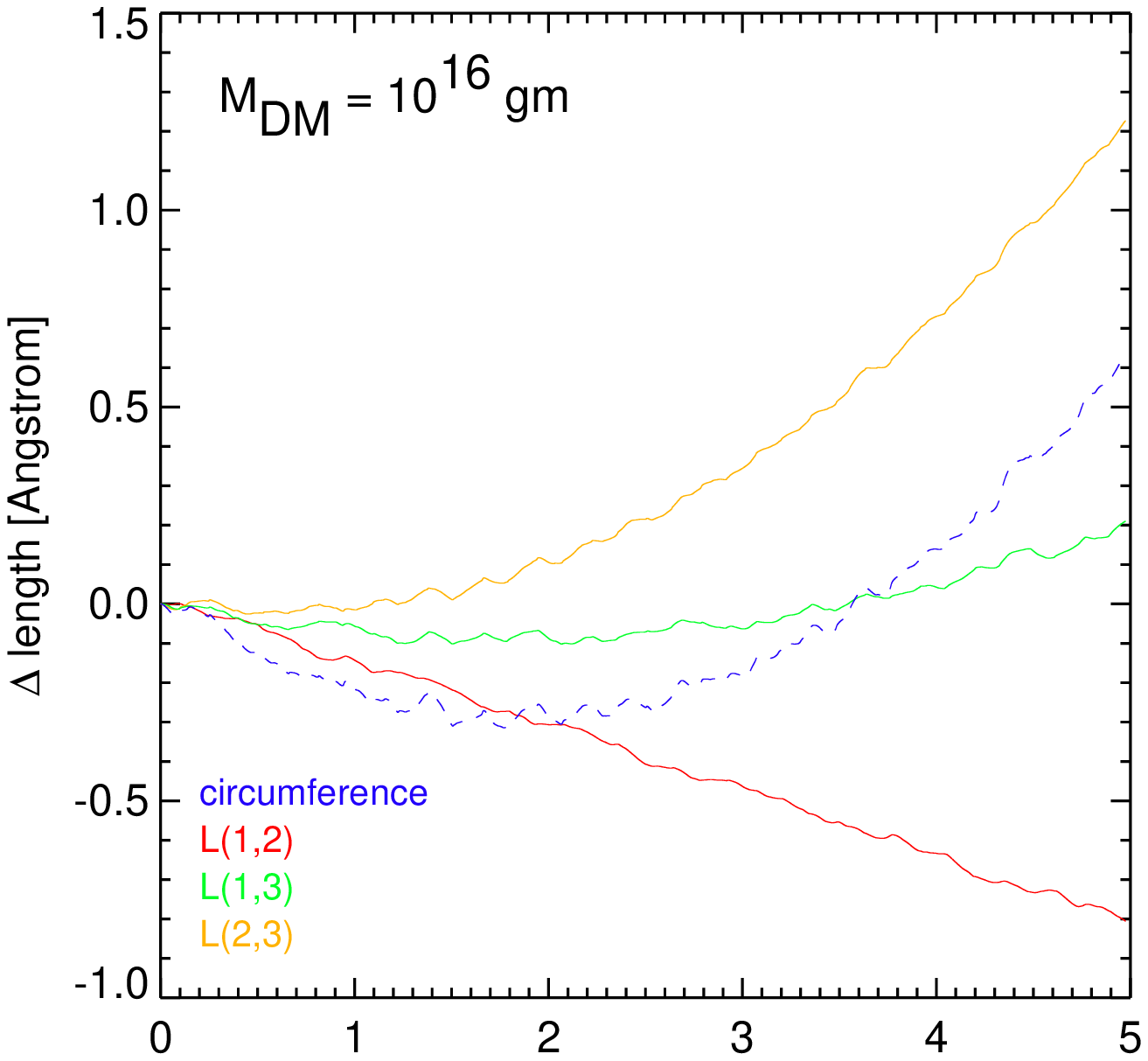,width=3.2in,angle=0}
}
\caption[]{\small
  (top) Quasi-Brownian walk of a single station from Keplerian orbit,
  perturbed only by DM objects of mass $M_{\rm DM} = 10^{17}$ (left)
  and $M_{\rm DM} = 10^{16}$ (right) over five years. The filled
  circles are placed at 3 month intervals. As expected, the total
  scale of the motion depends linearly on the dark matter
  mass. (bottom) Distance between stations ($L[i,j]$) as a function of
  time, significantly smaller than the bulk motion of the individual
  stations. The apparent periodicity in the impulse signal is due to
  periodic boundary conditions in the DM simulation. }
\label{fig:walk}
\end{figure*}

As expected, the simulations reveal myriad sharp, manifestly impulsive
short-timescale perturbations to the inter-station distances. Figure
\ref{fig:walk} shows the result of the perturbations from Keplerian
orbit due to dark matter only.  While the perturbations on a single
station are large (indeed, millions of times larger than the 1 sec rms
accuracy of LISA), the resultant LISA signals, ie changes in the
armlengths due to tidal effects, are indeed quite small - though
certainly measurable.  This is an important point for future
experiments, to which we shall return below. Note that for larger
masses the perturbation scale is larger, but there are fewer impulse
``events'' on a single station.

Figure 2 demonstrates the vastly different scales of the
perturbations, while figure 3 shows that despite the presence of
strong perturbations from major bodies, the passage of DM objects can
be detected as short time scale impulsive perturbations on the
inter-station timeseries at magnitudes significantly larger than the 1
sec rms accuracy of LISA. For instance, in the $M_{\rm DM} =
10^{18}$\,gm simulation, we see tens of events with $\delta d > 1$
\AA\ over $\Delta t < 10$\,days (see Figure 3b). The rate at this
impulse scale decreases quickly for masses less than $\sim 10^{15}$\,
gm.

In figure \ref{fig:detect} we show the details of a single DM event
from the 10$^{14}$\,gm DM simulation.  There are several important
characteristics of these near-field events discussed above and
demonstrated in this figure. First, the event duration is
significantly shorter ($< 1$ week) than the timescale for
perturbations due to the Moon (and other major bodies). Second, the
circumference changes during the event. Third, the time lag between
features in the inter-station distances is less than $\sim$2 days,
comparable to the quantity $b/v$. Fourth, assuming the LISA
decorrelation time is a few days or longer (as expected), since the
rms errors for a 24 hr timespan is smaller than the event scale, such
events should easily recognizable by their significant departure from a
2nd order polynomial fit.

We consider the typical events produced by 10$^{14}$\,gm DM as the
lower bound on the detectability given the advertised LISA
sensitivities. There appears to be several tens of such events in our
simulation of a 5 year mission. The upper bound on the detectable mass
(around 10$^{20}$\,gm) is set by two considerations: a) the event
rates drop at higher masses and b) the timescale for DM deflection
becomes comparable to timescale for the lunar perturbations. Indeed,
the maximum impact parameter for all masses, regardless of the actual
noise characteristics of LISA, is roughly set by $l_{\rm max} \approx
v_{\rm DM}\, \times 28$ days $\approx$ 3.7 AU $v_{230}$.

\section{Non-dark Impulsive Backgrounds}

LISA will also scatter inelastically off comets, asteroids and other
minor Solar System baryonic bodies \citep{ben+98}.  The pre-Phase A
study concluded that only the most massive asteroids produce
signatures with S/N $>$ 3 using the acceleration power-spectrum
analysis, with a rate of $\sim$ 0.08 events/yr, and are thus not a
significant background for gravitational wave detection. As discussed
above, we believe the detection rate will be much higher (by several
orders of magnitude) when searching for the impulsive signature; these
baryonic objects thus form an important background for our DM search.
However, even when the mass of such bodies are comparable to the dark
matter masses to which LISA is sensitive, at least two important
differences aid in distinguishing the signals in the LISA data stream.
First, the heliocentric velocities of the non-dark matter,
gravitationally bound in the Solar System, will be generally smaller
(by factors of several) than the dark population --- this has the
effect that the correlated peaks should have time-lags $\sim$60 hours
for sun-orbiting objects, rather than $\sim$6 hr for Galactic
DM. Second, and by far most important: non-dark matter reflects Sun
light, so removing this background reduces to, literally, {\it
looking} at it.

To make sure this is feasible, we estimate the brightness of such
light-reflecting objects. The characteristic size of an asteroid with
mass $M$ and average density $\bar\rho$ is,
$$R_c = 0.5~{\rm km}~\left(\frac{M}{10^{15}~{\rm gm}}\right)^{1/3} \
\left(\frac{\bar\rho}{{\rm 3~gm~cm}^{-3}}\right)^{-1/3}$$
The object size is $\approx 0.6$\,km for $\bar\rho = 1$ gm\,cm$^{-3}$,
appropriate for icy comets. If the object passes near LISA, then the
bolometric flux of reflected light at Earth is,
\begin{eqnarray}
f_\Earth & \approx & \frac{L_\sun R_c^2\,r}
        {16 \pi\, d^2_{{\rm LISA}-\Earth} d^{2}_{{\rm LISA}-\Sun}} \\
        & \approx & 7 \times 10^{-11}~~{\rm erg~s}^{-1},
\end{eqnarray}
where $L_\sun$ is the solar luminosity, a geometric/microphysical
reflectance of $r \approx 0.1$, the LISA--Earth distance $d_{{\rm
LISA}-\Earth} = 1$ AU $\times \sin 20^\circ/\sin 80^\circ = 5.2 \times
10^{12}$\,cm, and the LISA--Sun distance of $d_{{\rm LISA}-\Sun} = 1$
AU. The flux will be radiated mostly at optical and infrared
wavelengths. Assuming $\sim$1/3 (the precise number is not important)
of the bolometric flux is radiated in the $R$-band filter ($\lambda_c
\sim 6000$ \AA), the flux density in the $R$-band filter will be
$f_\nu \approx 6 \times 10^{-26}$ erg s$^{-1}$ Hz$^{-1} = 6$\,mJy or
$R \approx 14.5$\,mag.  These magnitude levels are routinely reached
by small aperture, large field-of-view telescopes.  LISA will subtend
a field--of--view of no more than 6$^\circ$ in diameter, and always
reside 80$^\circ$ from the Sun, observable for several hours per
night. The next generation of near-earth asteroid surveys (e.g.,
PanSTARRS and LSST, starting in 2006) will patrol the sky $\sim$10
magnitudes fainter, cataloging asteroids at Earth-LISA distances which
are smaller in size by a factor of $\sim$100, or down to masses of
$\approx 10^{9}$ gm. If there is sufficient interest in understanding
the local LISA gravitational background, the community might consider
the benefits of launching a small ($\sim 1$-meter) optical telescope
satellite in low-Earth orbit dedicated to observing the LISA field. If
a reflecting object is found in monitoring, once the ephemeris is
established and mass estimated, the effects upon the LISA baseline
evolution can be calculated, removing the non-dark signal.

\section{Next-Generation Experiments}

One of LISA's most obvious limitations as a DM detector is its
measurement sensitivity, any improvement in which would increase the
event rate significantly in both acceleration and impulse methods of
detection.  Another limiting factor is the correlation time;
lengthening this would similarly improve detection rates in the
impulsive channel.  Of course, since these are the main limitations
for LISA's gravity wave mission, they are already on the cutting-edge
of feasibility, and so these are difficult  levers to tweak.

A major limitation comes from the fact that LISA measures only tidal
forces on the constellation, significantly decreasing the power in the
signal.  To get a sense for what an enormous limitation this is,
consider Figure 1. The average deformation of the orbit of the
constellation is $\sim$four orders of magnitude larger than the tidal
displacement.  Increasing the LISA arm-length will thus increase
significantly the event rate.  Since collisions with larger impact
parameters have larger timescales, the finite correlation time tempers
this improvement for large arm-lengths; nonetheless, this would lead
to a significant improvement in resolution.  This provides additional
motivation for the construction of next-generation
laser-interferometric gravitational observatories with larger
baselines.  Another glaring limitation of LISA is the cacophonous
background of the visible solar system, whose signal on LISA is over
16 orders of magnitude larger than the signal we would like to
measure.  Moving this experiment, or a similar interferometer, outside
the solar system would be a tremendous improvement (if technologically
daunting). Incorporating such improvements into an experimental design
is beyond the scope of this letter; we will return to this issue in
future work.

\bigskip
\bigskip

{\it Note added in manuscript}: As this {\it Letter} was being
completed, we became aware of another group \citep{sc04} who
anticipated some features of the approach detailed above.
%
%
%
%

\acknowledgements

We would like to thank Nima Arkani-Hamed, Savas Dimopoulos, Matt
Holman, Scott Hughes, Stephen Shenker, Matt Strassler and Matias
Zaldarriaga for fruitful conversations. A.W.A. and J.S.B. are
supported by Junior Fellowships from the Harvard Society of Fellows.
This work was also supported by a research grant from the
Harvard-Smithsonian Center for Astrophysics (J.S.B.) and a visiting
position at the Rutgers High Energy Theory Center (A.W.A.).


\begin{figure*}[tp]
\centerline{\psfig{file=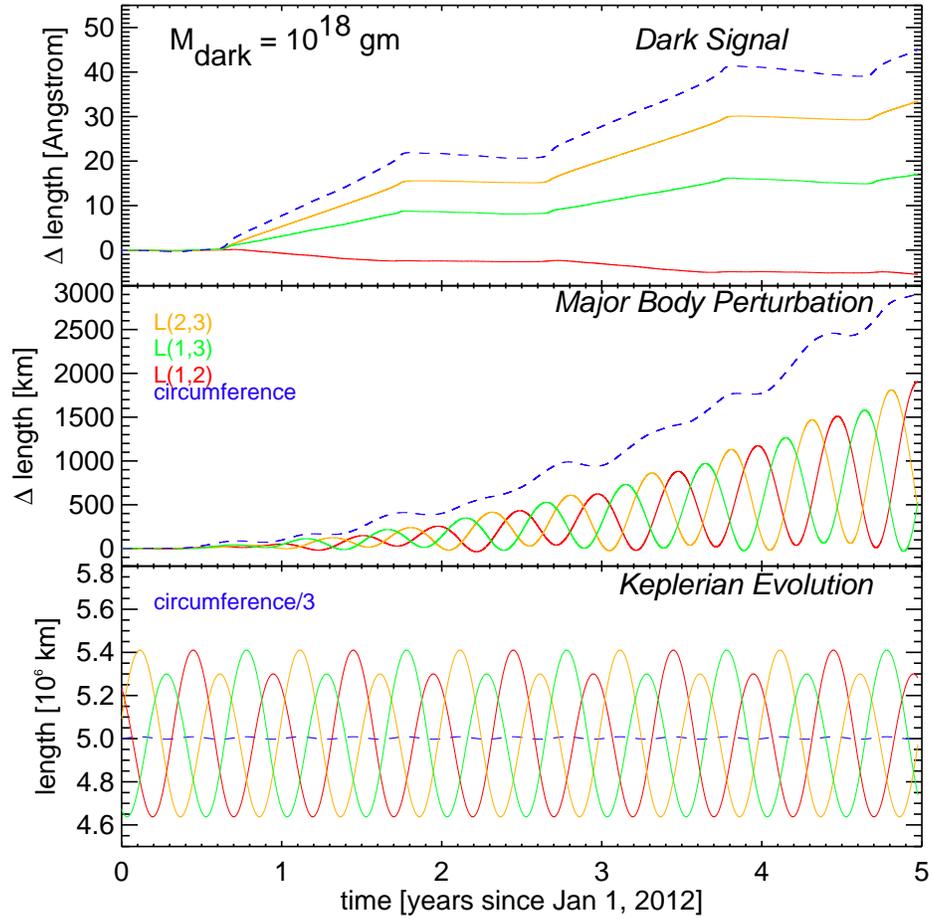,width=5in,angle=0}}
\caption[]{\small
  Simulation of inter-station distances in the presence of $M_{\rm DM}
  = 10^{18}$\, gm dark matter, including the orbit of the LISA
  constellation and perturbations from all inner solar system major
  bodies. The vast range of scales inherent in this dark matter
  detection technique is readily apparent.}
\label{fig:simulation}
\end{figure*}

\begin{figure*}[tp]
\centerline{\psfig{file=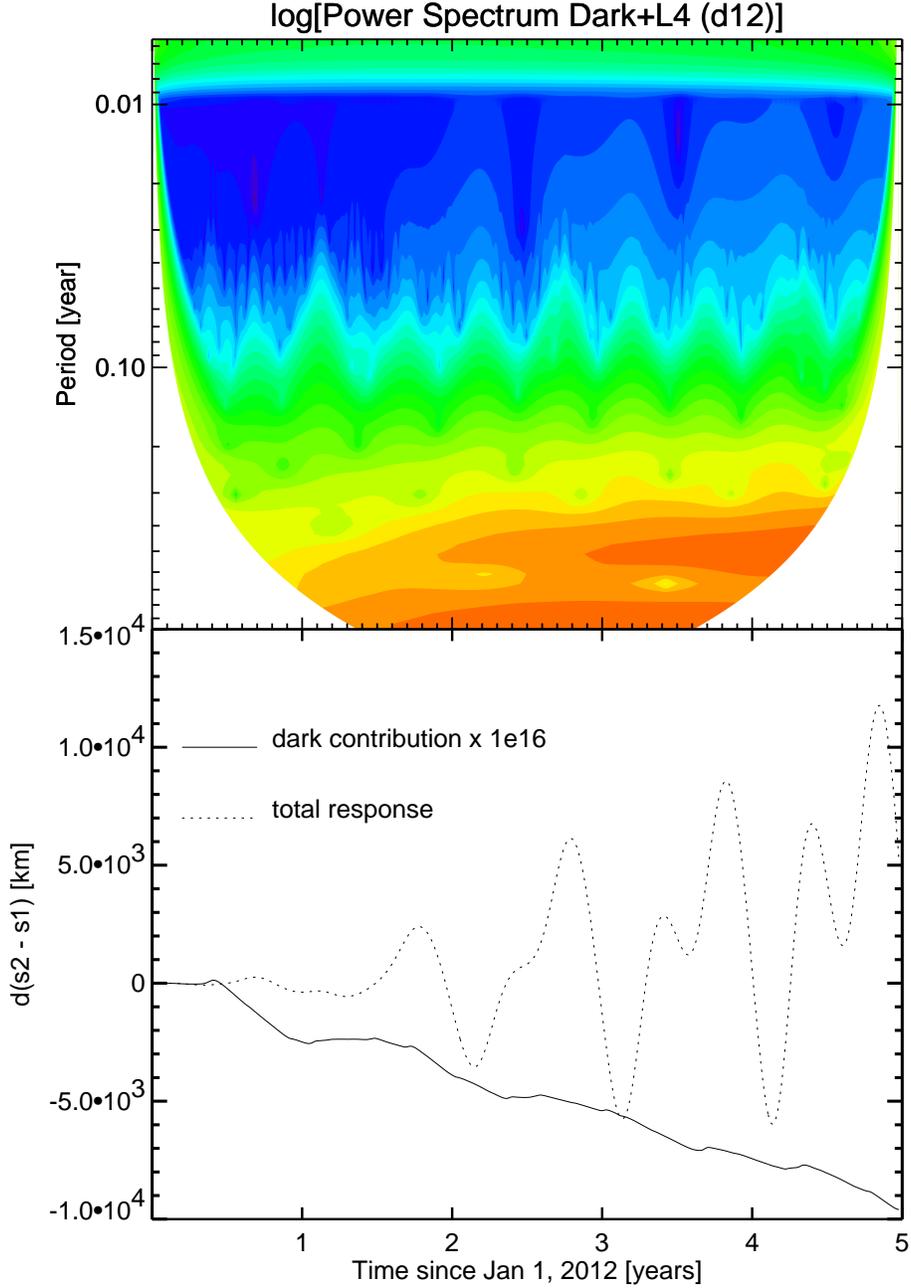,width=5in,angle=0}}
\caption[]{\small
	(top left) Time evolution of the power-spectrum of
	inter-station distances for the LISA orbit, determined by
	wavelet transform. Relative power increases from blue to
	red. Orbital perturbations due to all major Solar System
	bodies are included. Earth, which dominates the perturbations,
	has been excluded to retain a dynamic range of the DM and major
	body perturbations less than 10$^{16}$, the numerical
	precision of double floating point numbers. The contribution
	from the Moon is manifested in the short period spikes every
	28 days. (bottom) The time evolution of the inter-station
	distance (dotted line) and the contribution of from the dark
	matter in the simulation. Here we simulated 35 DM objects
	with $M_{\rm DM} = 10^{18}$\,gm streaming in the inner 11 AU
	of the Solar System.}
\label{fig:wavelet1}
\end{figure*}

\begin{figure*}[tp]
\figurenum{3}
\centerline{\psfig{file=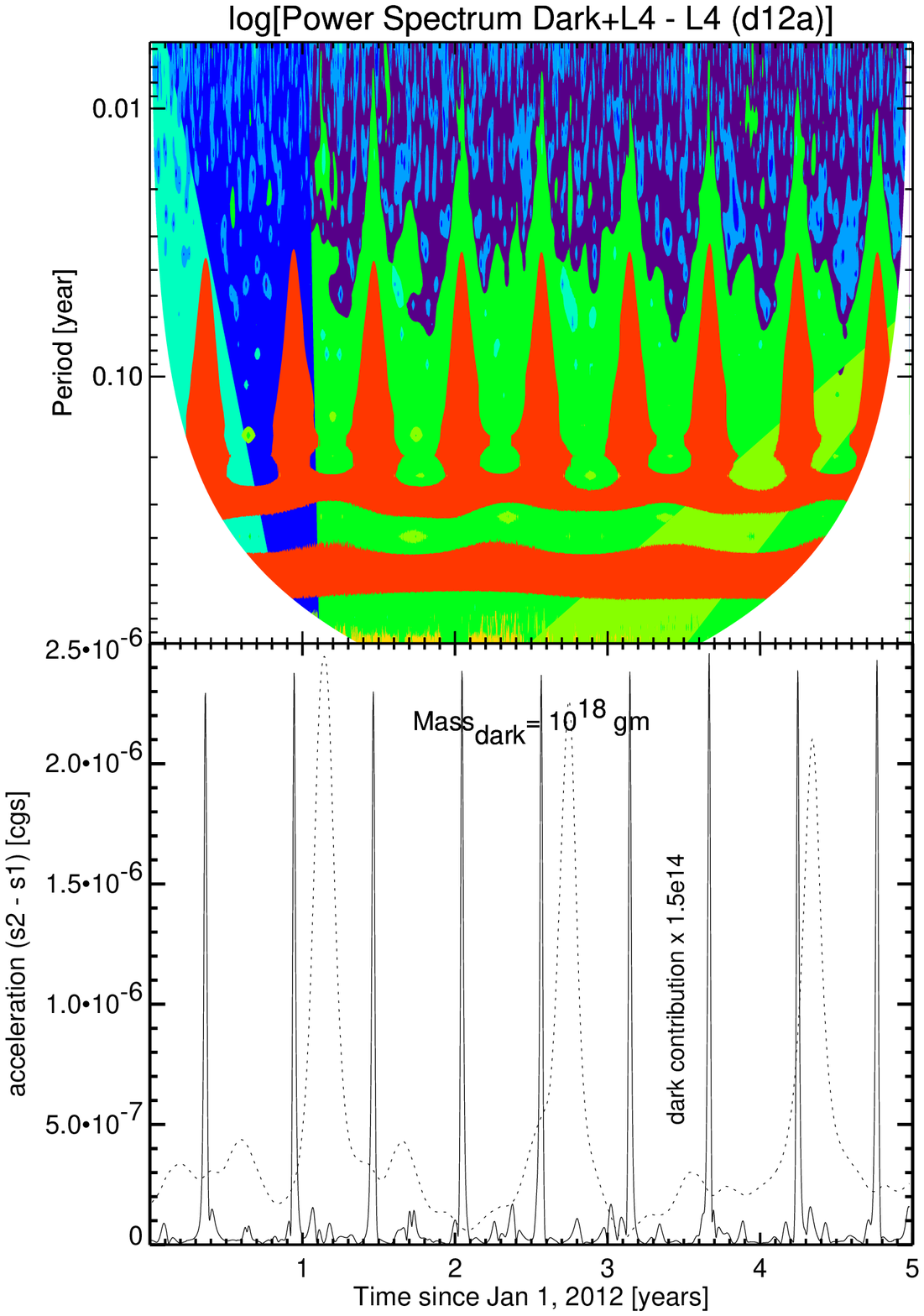,width=5in,angle=0}}
\caption[]{\small (continued)
	(top right) Demonstration that DM can be detected in the
	presence of significantly larger perturbations: time evolution
	of the power-spectrum of the difference between the
	inter-station acceleration with and without DM. The spikes from
	the DM signal are readily seen; the apparent periodicity is due
	to periodic boundary conditions in the simulation. The large
	blue cone in the first year is an artifact of the dynamic
	range in the relative precision on the two acceleration scales
	(bottom right) The time evolution of the total inter-station
	acceleration (dotted line) and the contribution of from the
	dark matter in the simulation (solid line).}
\label{fig:wavelet2}
\end{figure*}

\begin{figure*}[tp]
\centerline{\psfig{file=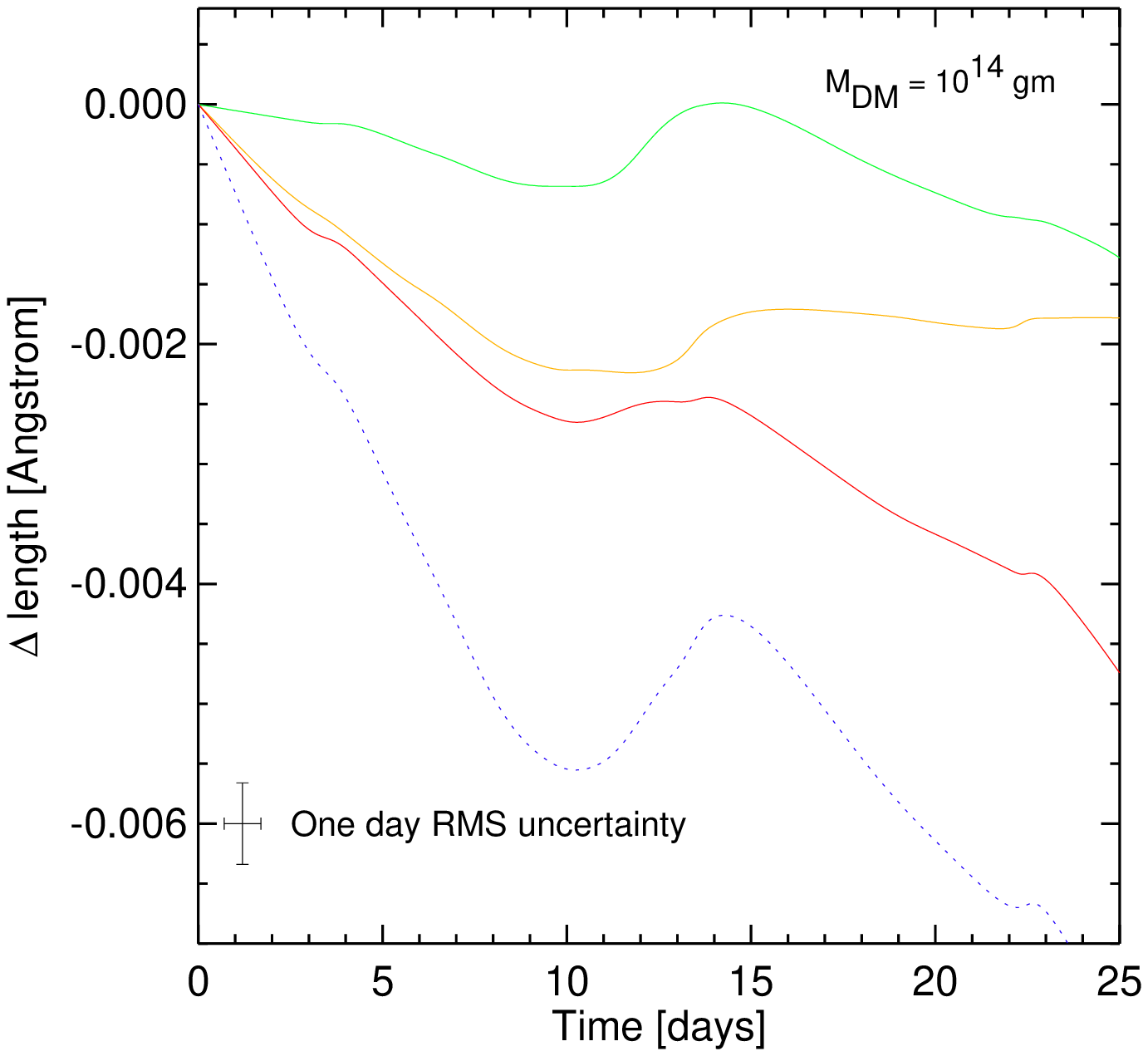,width=5in,angle=0}}
\caption[]{
        Zoom in on a generic DM passage event that occurred in a
        simulation with 2000 $M_{\rm DM} = 10^{14}$\,gm
        objects. Events generated by $M_{\rm DM} = 10^{14}$\,gm DM
        will produce detectable signals, if the decorrelation
        timescale is longer than several days (as currently
        anticipated).}
\label{fig:detect}
\end{figure*}

\end{document}